\documentclass{osa-article}
\journal{osajournal}
\articletype{Research Article}
\usepackage{amsmath}
\usepackage{lineno}
\usepackage{multicol}
\usepackage{multirow}
\usepackage{array}
    \newcolumntype{P}[1]{>{\centering\arraybackslash}p{#1}}
    \newcolumntype{M}[1]{>{\centering\arraybackslash}m{#1}}
\begin{document}

\title{Spatiotemporal pulse characterization with far-field beamlet cross-correlation}

\author{Slava Smartsev\authormark{1,$\dagger$,*}, Sheroy Tata\authormark{1,$\dagger$}, Aaron Liberman\authormark{1}, Michael Adelberg\authormark{1}, Arujash Mohanty\authormark{1}, Eitan Y. Levine\authormark{1}, Omri Seemann\authormark{1}, Yang Wan\authormark{1}, Eyal Kroupp\authormark{1}, Ronan Lahaye\authormark{2}, C\'edric Thaury\authormark{2} and Victor Malka\authormark{1}}

%\address{}

\email{\authormark{*}slava.smartsev@weizmann.ac.il} %% email address is required

\address{\authormark{1}Department of Physics of Complex Systems, Weizmann Institute of Science, Rehovot 7610001, Israel

\authormark{2}LOA, CNRS, \'Ecole Polytechnique, ENSTA Paris, Institut Polytechnique de Paris, 181 Chemin de la Huni\`ere et des Joncherettes, 91120 Palaiseau, France}

%%%%%%%%%%%%%%%%%%% abstract %%%%%%%%%%%%%%%%

\section*{Abstract}
We present a novel, straightforward method for spatiotemporal characterization of ultra-short laser pulses. The method employs far-field interferometry and inverse Fourier transform spectroscopy, built on the theoretical basis derived in this paper. It stands out in its simplicity: it requires few non-standard optical elements and simple analysis algorithms. This method was used to measure the space-time intensity of our 100 TW class laser and to test the efficacy of a refractive doublet as a suppressor of pulse front curvature (PFC). The measured low-order spatiotemporal couplings agreed with ray-tracing simulations. In addition, we demonstrate a one-shot measurement technique, derived from our central method, which allows for quick and precise alignment of the compressor by pulse front tilt (PFT) minimization and for optimal refractive doublet positioning for the suppression of PFC.

% \begin{abstract}

% \end{abstract}

%%%%%%%%%%%%%%%%%%%%%%%%%%  body  %%%%%%%%%%%%%%%%%%%%%%%%%%

\section{\label{sec:intro}Introduction}
For more than three decades, the use of chirped pulse amplification (CPA) has opened the frontier of high energy, ultra-short laser pulses\cite{STRICKLAND_OC_1985}. When used in concert with other laser amplifier technologies, scientists have demonstrated CPA laser chains with peak powers in the gigawatt and even petawatt ranges, and have set their sights towards realizing exawatt class lasers\cite{Danson_HPLSE_2019}. These intense, short laser pulses provide a powerful tool in a wide range of fields, and in particular in the study of laser-matter interaction in the relativistic regime\cite{mourou-rmp06}.

In addition to temporally stretching the pulse, CPA laser chains typically include beam expanders, used to limit the fluence during the amplification stage, which are essential to mitigate unwanted nonlinear effects and avoid damage to the optical components. It is often more practical to use refractive as opposed to reflective beam expanders. However, since a broadband spectrum is necessary in order to achieve an ultra-short pulse, expansion using refractive lenses has the side effect of accumulating chromatic aberrations. This can significantly deteriorate the spatiotemporal properties of the pulse at focus. Thus, the use of refractive beam expanders is often limited to only the early stages of amplification, before the beam's diameter grows too large, when these undesirable effects are minimized. Even with perfect alignment, a conventional beam expander that utilizes the same glass for both lenses introduces a specific spatiotemporal coupling (STC), known as pulse front curvature (PFC). The part of the beam that travels across the optical axis is delayed more than others since it accumulates a larger group delay. The relative delay scales quadratically with the beam size \cite{Bor_OL_1989,Bor_89}, which makes it especially relevant for large diameter high-intensity beams. This leads to a curved pulse front that is radially symmetrical across the beam. In addition to PFC, a spatially varying group delay dispersion (GDD) is introduced, broadening the pulse in time non-uniformly across the beam. This can be an important effect for pulses having a duration much shorter than 25 fs. Also, in real systems, due to imperfections in optical components and misalignment of beam expanders and stretcher-compressor pairs, other spatiotemporal couplings are introduced, such as pulse front tilt (PFT) or/and spatial chirp \cite{Jolly_IOP_2020}. The STCs mentioned above are the most dominant and are known as low order STCs. However, in real CPA laser systems more exotic types of STCs may arise. Among them are complex chromatic spatial phase aberrations induced by defects in the compressor optics and spatially non-uniform spectral composition due to pump depletion during amplification in a saturated regime\cite{Jolly_IOP_2020,jeandet2021survey}.

In most cases, any spatiotemporal couplings are undesirable because they increase pulse duration and reduce peak intensity and contrast at the focus \cite{Bourassin_OE_2011,Li_IOP_2017,Li_OE_2018}. However, in some cases, one can exploit spatiotemporal couplings in a controlled and intricate way to manipulate the dynamics of the intense pulses at the focal region \cite{Sainte_Marie_Optika_2017,Froula_NPH_2018}. Control over the velocities of energy depositions along the focal region through the sculpting of customized STCs enables novel experiments in Raman amplification and ionization waves \cite{Turnbull_PRL_Jan_2018,Turnbull_PRL_May_2018} and paves the way toward a new generation of laser-driven particle accelerators \cite{Debus_2019_PRX,Caizergues_NP_2020,Palastro_PRL_2020}. Techniques for accurate and straightforward measurement of STCs lie at the heart of our ability to mitigate unwanted couplings and enable the use of STCs as a critical degree of freedom in an experiment.

In parallel to state-of-the-art temporal metrology development \cite{Walmsley_09,Monmayrant_2010,Cai_2020}, significant progress has been made in measuring STC during the last decades. The development of spatio-temporal metrology is well summarized in reviews \cite{Akturk_2010,Dorrer_2019,Jolly_IOP_2020}, while the cutting edge work is presented in recent papers \cite{ Jean-Baptiste_20,Kim_21,Grace_2021}. Some of these state-of-the-art techniques provide the complete spatiotemporal characterization of near-visible femtosecond laser beams. The cost, in most cases, is complex optical setups and heavy post-processing algorithms, which makes it challenging to use them on a regular basis. Other STC measurement techniques sacrifice completeness for the sake of simplicity, and are typically limited to spatially varying pulse front delay measurement (first order spectral phase).

This article demonstrates a straightforward method to measure STC that is based on far-field interferometry and inverse Fourier transform spectroscopy. Put simply, we measure the beam's group delay and temporal width at a discrete set of points along the beam with different radial and angular coordinates. With simple optics and analysis, our approach can accurately ascertain the pulse front in time and space together with the spatially-varying pulse duration. We demonstrate the efficacy of this technique by characterizing the spatio-temporal intensity profile of the 25 fs 100 TW laser at Weizmann Insitute of Science (WIS) \cite{MRE_WIS}, and extracting the values of the PFT, PFC, and spatially-varying temporal duration. We show that for our system, higher order STCs have a negligible effect on the intensity. In addition, using this measurement technique, we investigate the performance of a refractive doublet, which we refer to as the "PFC compensator", as a means to suppress unwanted PFC.

\section{\label{sec:meas}Far field beamlet cross-correlation STC measurement}
The proposed method is self-referenced and based on linear field cross-correlation (also known as inverse Fourier transform spectroscopy) \cite{Bor_89,Miranda_2014,Pariente_NatPhot_2016}. Similar to other spatially resolved Fourier transform techniques\cite{Miranda_2014,Pariente_NatPhot_2016}, it uses a small diameter central portion of the beam as a reference (reference beamlet). However, instead of spatially expanding the reference beamlet to the full beam size and comparing temporally in the near field, our method "directly" compares the reference beamlet with test beamlets in the far-field, similar to far-field slits interferometry \cite{Netz_2000,Li_19}. It significantly reduces the complexity of the optical setup and alignment since it relies on focusing optics and focal spot imaging, which are usually already part of the optical setup. 

The far-field beamlet cross-correlation is performed as follows: first, the two near-field beamlets, selected by means of a special mask, are focused and form an interference pattern in the far-field. Second, the beamlet pair is cross-correlated by changing the delay of the reference beamlet by a piezo actuator attached to the central portion of the mirror. Finally, the fringe contrast envelope as a function of delay is used to compare the beamlets temporarily. The contrast envelope's maximum corresponds to zero group delay (synchronization) between the beamlets while the contrast width reveals the temporal broadening, which can be caused by the second-order relative spectral phase (GDD). Repeating this cross-correlation for test beamlets at different spatial locations reveals the STC for the entire beam. The advantage of the method is that the spatial resolution of the STC measurement could be easily adjusted if needed. Usually, there are low order STCs; therefore, total scanning time could be reduced by picking a proper spatial resolution of the measurement. 

This technique allows for the measurement of STC for beams with custom spatial beam shaping that result in intricate spatial aberrations. It is well suited for the direct measurement of the group velocity of the intensity peak for long focal depth optics, such as axiparabola experiments, which have a large amount of controlled spherical aberration \cite{Smartsev_19,oubrerie_2021}.

\begin{figure}[t]
		\centering
		\includegraphics[width=13cm]{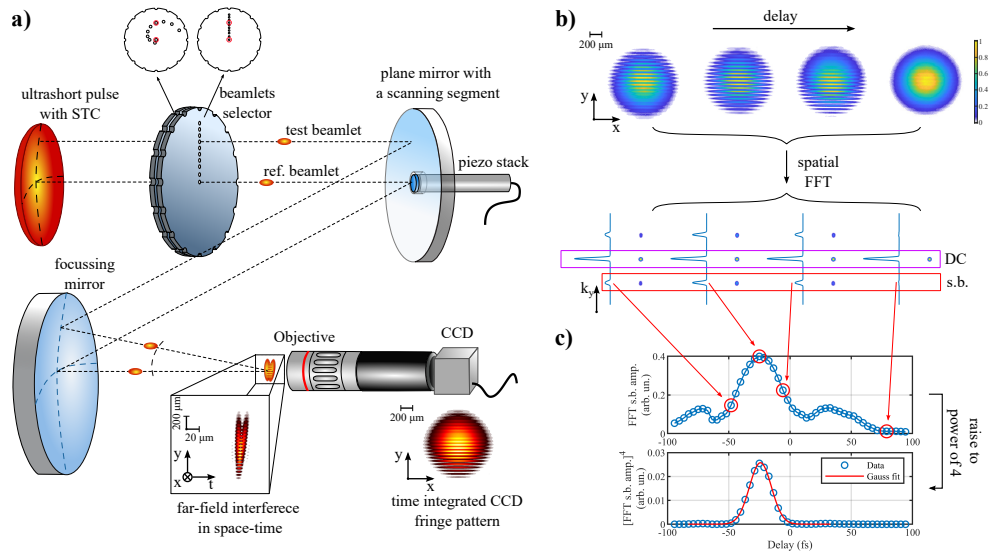}
		\caption{\label{fig:setup} \small {Simplified setup and analysis flow. \textbf{a)} Large diameter ultrashort laser pulse with STC impinges on beamlet selector, which splits the large beam into two small beamlets: test and reference. The reference beamlet stays on an optical axis and is delayed with a mirror attached to a piezo stack. Both beamlets are focused with a parabolic mirror, imaged by a microscope objective, and interfere on the CCD sensor. \textbf{(b)} Selected experimental fringe patterns for different delays of one beamlet pair. \textbf{(c)} (top) Fringe visibility as a function of delay is calculated using spatial FFT through the value of the side-band (s.b.) with respect to the DC value; (bottom) the side-band value is raised to the fourth power and fitted with a Gaussian.}}
\end{figure}

\section{\label{sec:setup}Experimental setup}
We measured the STC of the HIGGINS Ti:sapphire 2x100 TW laser system at the WIS, which delivers two 25 fs pulses with an energy up to 2.3 J per beam at a repetition rate of 1~Hz and a beam diameter of 60 mm in transport. The spectrum of the laser has a double peak shape (see Fig.~\ref{fig:spectra}) and is centered around 800 nm (or 375 THz in frequency), with a spectral width of 40 nm at FWHM.
We measured the STC of both laser beams and obtained similar results; therefore, we will present results for one beam (B1). The simplified experimental setup used for the beamlet cross-correlation STC measurement is depicted in Fig.~\ref{fig:setup} (a). Fully amplified, attenuated, and compressed laser pulses impinge on the beamlet selector mask, which consists of two perforated disks. The orientation of the discs allows for the selection of the test beamlet from a discrete set of radial and angular positions while the reference beamlet passes through the central hole (see Appendix~\ref{appendix:beamlets_selector}). The 2.65 mm perforations were selected to minimize diffraction effects and to have approximate uniformity of STC across the beamlet.

In the particular case shown in Fig.~\ref{fig:setup} (a) the test beamlet is 27 mm apart from the reference beamlet. After the selector, the two beamlets are reflected at a 45 degree angle from a specially designed segmented plane mirror. This mirror has a cutout in the center in which a small movable mirror attached to a piezo actuator is installed (see Appendix~\ref{appendix:delay_mirror}). The reference beamlet is reflected from this actuated mirror, allowing for precise control over the delay between the beamlets. The delay scales with the added length to the optical pathway of the reference beam, with a geometric factor that arises from the angles of reflection (see Appendix~\ref{appendix:delay_mirror}). Then, the beamlets are focused with a 2 m focal length off-axis parabola (f/33), and the far-field is imaged to a 16-bit CCD camera with a microscope objective.

The interference patterns, shown in Fig.~\ref{fig:setup} (a), correspond to a 25 fs delay between the beamlets. Since the test beamlet has a slightly different k-vector and is delayed with respect to the reference, the interference pattern is not symmetric. More visible fringes appear at the region of better overlap in space-time. The space-time interference is shown for the case when the two beamlets are perfectly compressed (no global GDD). However, the time-integrated fringe pattern will not change in the case of global GDD being applied on both beamlets. For each beamlet pair, a set of measurements with varying delays is taken and the process is repeated for different test beamlets. An example of this data analysis sequence is shown in Fig.~\ref{fig:setup} (b). First, interference signals are processed to remove the background and crop the image to keep the main lobe; then, a standard spatial FFT is applied, yielding a signal with a DC term and two sidebands. The highest overall contrast of the fringe pattern corresponds to the highest value of the sideband. The relative amplitude of the sideband with respect to the DC peak is extracted and plotted as a function of delay in Fig.~\ref{fig:setup} (c)(top). This plot represents the beamlets' cross-correlation envelope. Finally, we raise the cross-correlation to the fourth power to match the width of the transform limited pulse as shown in Appendix~\ref{appendix:theory} and fit it to a Gaussian. The Gaussian fit parameters such as delay and width provide the relative first and second-order spectral phase of the test beamlet with respect to the reference; from these measurements, the pulse front and temporal width of the intensity along the beam is reconstructed. 

\section{\label{sec:results}Experimental results and discussion}
In the experiment, we measured the STC of one of our laser beams (B1) in two cases. The first is with the STC that results from our laser chain and the second is with the addition of a specially designed doublet lens to correct laser chain accumulated PFC \cite{Jolly_doub_20}. The doublet, which is placed in the middle of the final beam expander, allows for the tuning of PFC by adjusting its position in the diverging beam that is in the expander \cite{Kabacinski_2021}. For more details about the doublet we use, please see Appendix~\ref{appendix:laser_chain}.

\begin{figure}[ht]
		\centering
		\includegraphics[width=13cm]{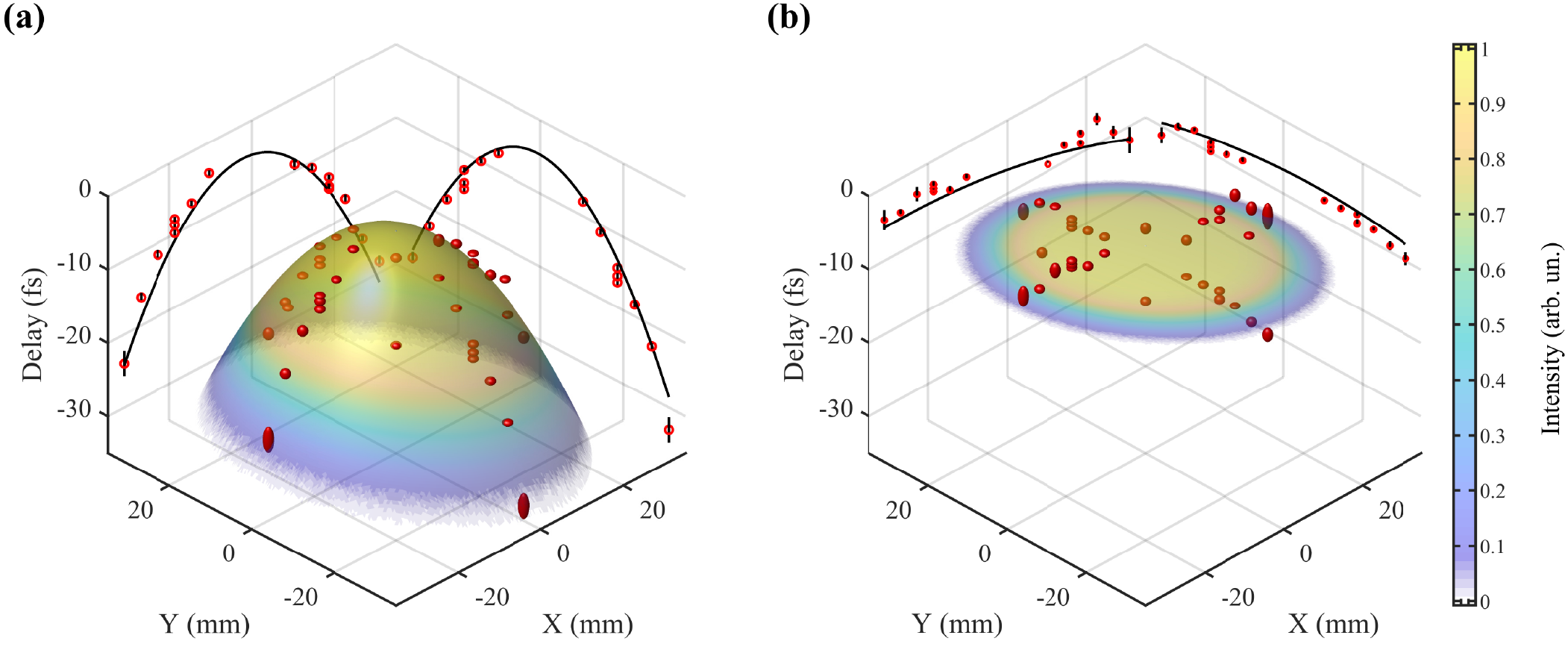}
		\caption{\label{fig:stc_delay} \small {Measured pulse front delay across the beam and corresponding semi-transparent fitted surfaces. (a) Uncorrected case and (b) Corrected case with the doublet. The beam intensity is encoded in the color of the surface. The x-y cross-sections of data points and surfaces are shown in corresponding planes.}}
\end{figure}

\begin{figure}[t]
		\centering
		
		\includegraphics[trim={0.01cm 0 0.01cm 0},clip,width=0.7\textwidth]{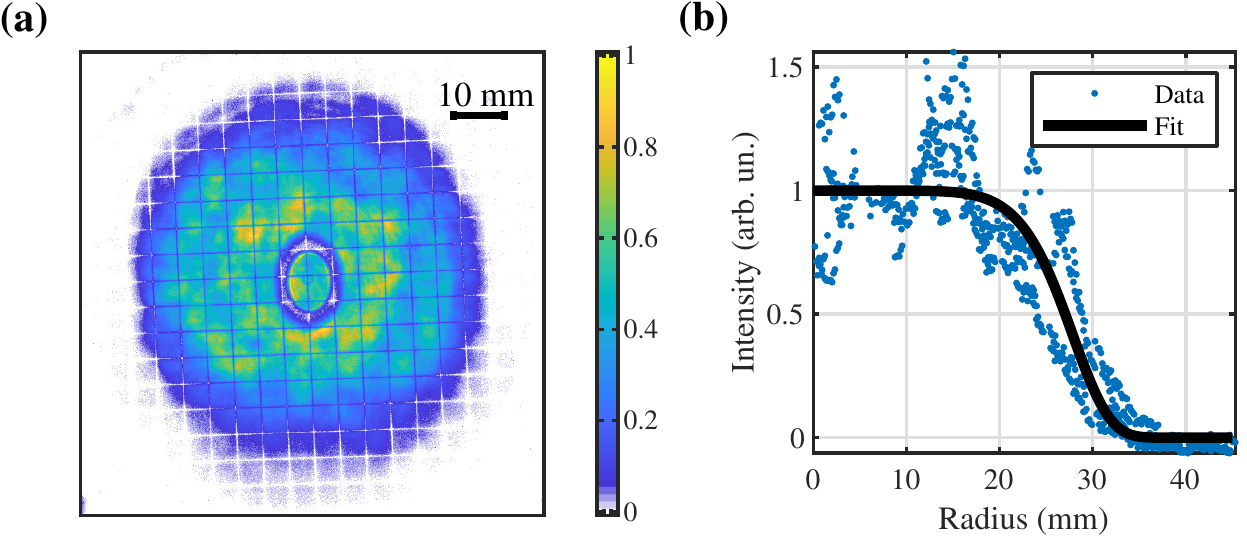}
		\caption{\label{fig:near_field} \small(a) Near field intensity profile. (b) Radial fit for intensity X-Y cross-sections. }
\end{figure}

We performed beamlet cross-correlation across the X-Y plane of the beam in 4 mm steps and took four additional points between the X-Y axes in the near field.
As we have shown analytically in Appendix \ref{appendix:theory} Eq.~\ref{eq19} a Gaussian approximation of the beamlet cross-correlation raised to the fourth power has the functional form:
\begin{equation} \label{eq1_main}
\exp\left(-\frac{1}{2\sigma_{bc4}^2}\tau_1^2\right)
\end{equation}
where $\tau _1$ is the group delay between the beamlets and $\sigma_{bc4}$ is the width of the correlation raised to the fourth power. We plot the fit extracted $\tau _1$ as a function of the test beamlet near-field position X-Y in Fig.~\ref{fig:stc_delay} for both the uncorrected and corrected PFC cases. The vertical size of the data points (or black vertical solid lines on x-y cross-sections) are the 95\% confidence bounds of the fit. To test the stability of our measurement, we compared data for the |x|,|y|=20 mm points from the beginning, middle, and end of the scan. As can be seen, these data points nearly coincide within the measurement error. Next we fit the pulse front delay data to low order polynomials to extract the corresponding PFT and PFC for both cases. The fit function was $\tau_1(x,y)=\tau_0+\mbox{PFT}_xx+\mbox{PFT}_yy+\mbox{PFC}(x^2+y^2)$. The fit surface is plotted with the data points, its color encoded with radially averaged intensity (see Fig.~\ref{fig:near_field}). As can be seen in Table~\ref{table:PFT_PFC}, our laser pulse front is dominated by PFC (no doublet case), which is to be expected since we used only refractive telescopes before compression. It can also be seen that the corrective doublet successfully suppressed the PFC by more than one order of magnitude.

It is worth noting that before this experimental run, the PFT and the PFC of the corrected case were optimized by grating alignment and doublet longitudinal positioning, respectively, using a fast measurement procedure based on beamlet interference. This will be discussed in detail later.

\begin{table}[ht]
\renewcommand{\arraystretch}{2}
\footnotesize
\centering
\begin{tabular}{ |P{2cm}||P{2cm}|P{2cm}| }

 \hline
 &No doublet& With doublet \\
 \hline\hline
$\mbox{PFT}_{\mbox{x}}~\mbox{(fs/mm)}$ &$-0.114\pm0.048$& $-0.102\pm0.039$\\
 \hline
$\mbox{PFT}_{\mbox{y}}~\mbox{(fs/mm)}$&$0.027\pm0.048$& $-0.029\pm0.039$\\
 \hline
$\mbox{PFC (fs/mm$^2$)}$&$-0.0239\pm0.0028$& $-0.0019\pm0.0023$\\
 \hline
\end{tabular}
\caption{\label{table:PFT_PFC} \small {Extracted fit parameters for low-order pulse front delay for uncorrected and corrected cases.}}
\end{table}

The next step is to evaluate the spatially-varying temporal width. For this purpose we utilize the beamlet cross-correlation width $\sigma_{bc4}$. As we have shown in Appendix~\ref{appendix:theory}, Eq.~\ref{app:eq20} the width is composed of three terms: 

\begin{equation} \label{main:eq2}
\sigma_{bc4}^2(x,y)=\frac{1}{2c_B^2}+\frac{x^2+y^2}{8\sigma_0^2\omega_c^2}+\frac{c_B^2\beta_r (x,y)^2}{8}
\end{equation}
The first term is the transform-limited width which is inversely proportional to the spectral bandwidth $c_B$. The second is a geometric term that broadens the cross-correlation as the distance $\sqrt{x^2+y^2}$ between the beamlets in the near-field grows and is inversely proportional to the size of the beamlets in the near-field $\sigma_0$ and the central laser frequency $\omega_c$. This term stems for the fact that the k-vectors of the beamlets differ because of the distance between them in the near-field. Finally, the third term stretches the cross-correlation when the test beamlet broadens relative to the reference beamlet. We shall assume that the relative GDD $\beta_r(x,y)$ is the cross-correlation broadening source. In general, it could include relative spectral narrowing and/or relative spectral phase of higher order than GDD. The measurement is insensitive to global GDD and thus GDD absence needed to be independently verified by Wizzler measurements (shown in Appendix~\ref{appendix:BCC_spectrum}) at the central part of the beam in near-field as required by inverse Fourier spectroscopy for full space-time reconstruction.

\begin{figure}[ht]
		\centering
		\includegraphics[width=13cm]{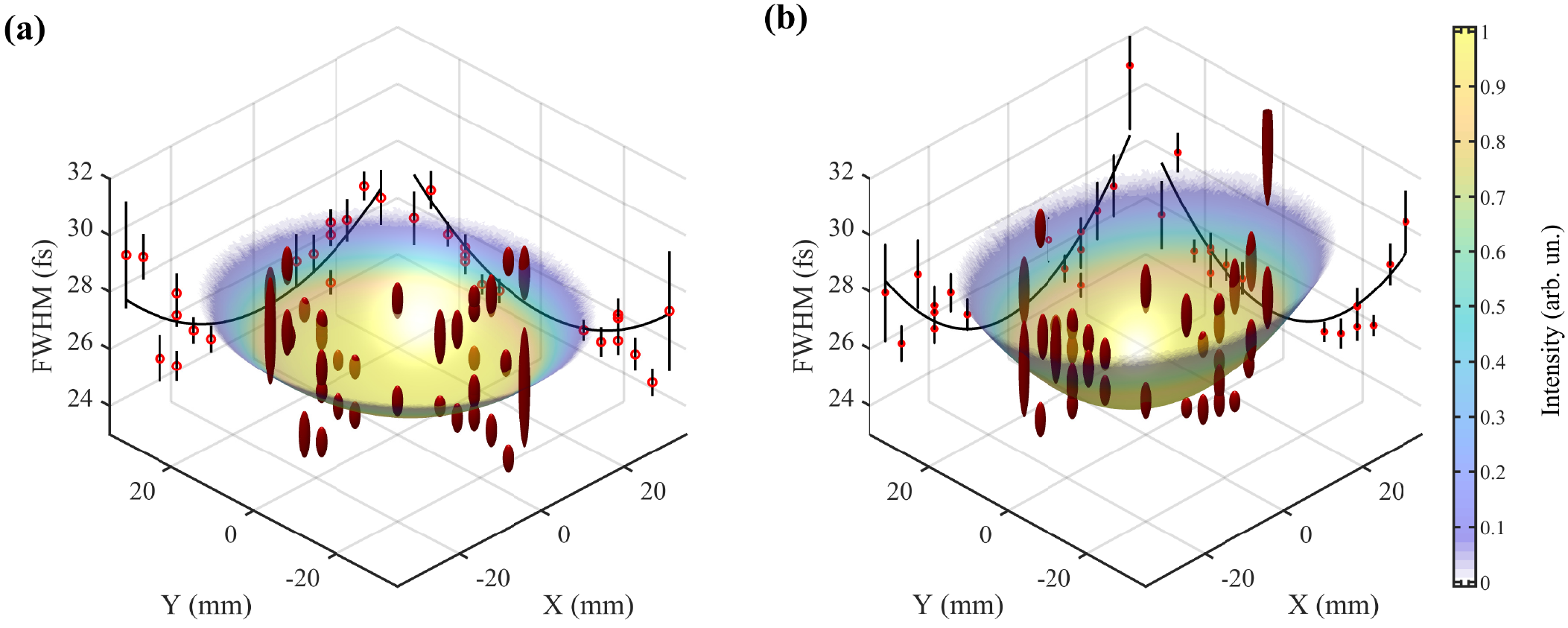}
		\caption{\label{fig:stc_width} \small {Measured near-field intensity temporal width across the beam and corresponding semi-transparent fitted surfaces. (a) Uncorrected case and (b) Corrected case with the doublet. The beam intensity is encoded in the color of the surface. The x-y cross-sections of data points and surfaces are shown in corresponding planes.}}
\end{figure}

Our laser has a double peak rather than a Gaussian spectrum; therefore the beamlets' cross-correlation width $\sigma_{cb4}$ is narrower than the temporal intensity width by a factor of $d=1.27$ as we show in Appendix~\ref{appendix:BCC_spectrum}. The final spatially varying temporal width is obtained by removing the geometrical term in Eq.~\ref{main:eq2} (using $\omega_c=2.36\cdot10^{15}$ rad/s, and $\sigma_0=1.1$ mm, which corresponds to the beamlet FWHM in the near field: $FWHM_0=2.65$ mm) and re-scaling by the deconvolution factor $d$. We plot intensity temporal width (FWHM) as a function of the test beamlet position X-Y in Fig.~\ref{fig:stc_width} for PFC uncorrected and corrected cases. The vertical size of the data points (or black vertical solid lines on x-y cross-sections) are the 95\% confidence bounds of the fit. Then we fit the data to low order polynomials $\mbox{FWHM}(x,y)=\mbox{FWHM}_0+\mbox{L}_xx+\mbox{L}_yy+\mbox{Q}(x^2+y^2)$, the coefficients of which are provided in Table~\ref{table:width_fit}.

\begin{table}[ht]
\renewcommand{\arraystretch}{2}
\footnotesize
\centering
\begin{tabular}{ |P{2cm}||P{2cm}|P{2cm}| }

 \hline
 &No doublet& With doublet \\
 \hline\hline
 $\mbox{FWHM}_{\mbox{0}}~\mbox{(fs)}$ &$24.6\pm0.64$& $24.46\pm0.68$\\
 \hline
$\mbox{L}_{\mbox{x}}~\mbox{(fs/mm)}$ &$-0.013\pm0.022$& $0.0063\pm0.0231$\\
 \hline
$\mbox{L}_{\mbox{y}}~\mbox{(fs/mm)}$&$0.0031\pm0.022$& $-0.0247\pm0.0231$\\
 \hline
$\mbox{Q (fs/mm$^2$)}$&$0.0025\pm0.0013$& $0.0040\pm0.0014$\\
 \hline
\end{tabular}
\caption{\label{table:width_fit} \small {Extracted fit parameters for pulse width for uncorrected and corrected cases.}}
\end{table}

From the results depicted in Table~\ref{table:width_fit} we can see that in both uncorrected and corrected cases, the beam has a transform-limited width of approximately 25 fs at the center, which assumes perfect compression verified by independent Wizzler measurements. The width grows slightly from the center toward the beam's edge. For example - using only quadratic slope - at a distance $r=30$ mm from the center, the width grows by 2 and 4 fs for the uncorrected and corrected cases, respectively. This, in turn, implies a temporal broadening corresponding to a relative GDD of 95 and 135 fs$^2$. These relative GDD values are higher than predicted from the ray-tracing simulation (see Appendix~\ref{appendix:laser_chain}); however, additional broadening could be explained by spectral narrowing toward the beam's edge.
 
For each case (uncorrected and doublet corrected), we scanned 36 pairs of beamlets with 50 delay steps at a 1 Hz rate. Beamlet switching took no more than 10 seconds; therefore, the total scan time was around 36 mins. The total data processing using a desktop computer with a dual Intel Xeon processor @ 2.2GHz and 64Gb RAM took 13 minutes. As can be seen, we sampled much more than needed for the low order fit we used in each scan; both in temporal - 50 delay steps - and in spatial - 36 beamlets pairs - resolution. The total time with optimized sampling and optimized spatial resolution could be reduced for the same final result.

\begin{figure}[t]
		\centering
		\includegraphics[width=13cm]{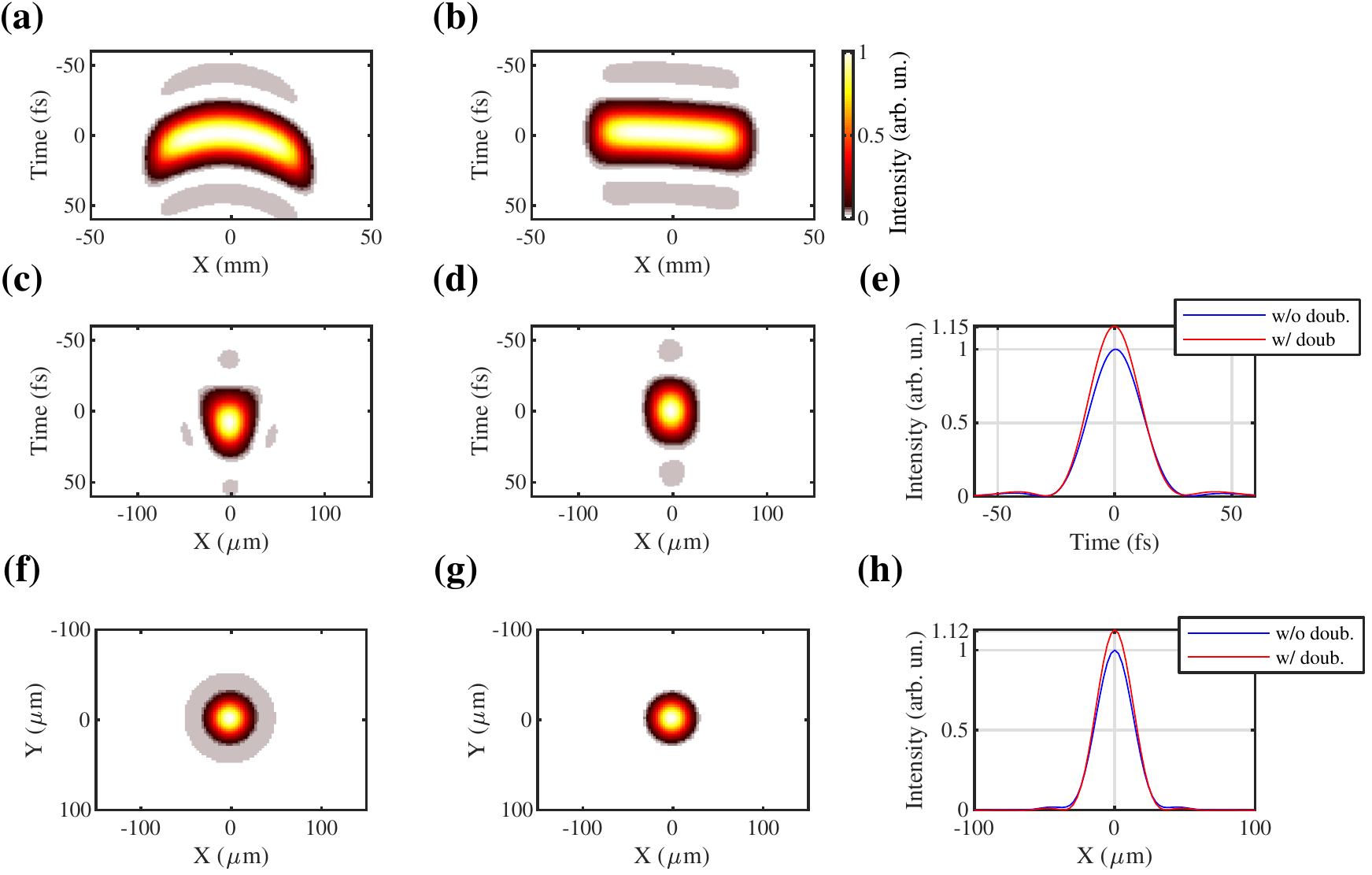}
		\caption{\label{fig:sim_far_field} \small Simulated spatiotemporal intensity cross-sections at near (a-b) and far-field (c-d) based on measured STC parameters, with time-integrated far-field intensities (f-g). Columns (a,c,f) and (b,d,g) correspond to uncorrected and doublet corrected cases, respectively. (e) far-field intensity along the time axis. (h) far-field time-integrated intensity along the X-axis.}
\end{figure}

Next, we used the measured STC parameters to evaluate spatiotemporal intensity at the far-field by numerical simulation. For simplicity, we used relative GDD as the only broadening source estimated from the quadratic slopes in Table~\ref{table:width_fit}. Cross-sections of the simulated space-time intensity profiles are shown in Fig.~\ref{fig:sim_far_field}; Columns (a,c,f) and (b,d,g) correspond to doublet uncorrected and corrected cases, with row (a,b) corresponding to near-field, row (c,d,e) to the far-field spatiotemporal intensity, and row (f,g,h) corresponding to time-integrated intensity. As can be seen in the near-field intensity cross-section along the optical axis in (e), the pulse front curvature, which is dominant in the uncorrected case, reduced the peak intensity by a factor of 1.15 from the corrected case. The temporal width of the uncorrected case was larger by a factor of 1.04. Thus, as can be seen from time-integrated far-field intensity cross-sections (h), the larger PFC term in the uncorrected case reduced the focal spot intensity of the beam by a factor of 1.12. 

As mentioned above, prior to data gathering with the scanning STC measurement method described in this paper, a one-shot measurement technique was used to align the optics to minimize PFT and to eliminate PFC in the corrected PFC case. We demonstrate this technique for the case of fast PFT measurement. A similar measurement of PFC would differ only in the mask used. This one-shot technique is based on beamlet delay which is extracted directly from the internal structure of the far-field interference (similar to far-field slits interferometry\cite{Netz_2000,Li_19}). For this, we use a modified beamlets selector ("PFT tool"), a mask which selects two symmetrical off-axis beamlets (see Appendix~\ref{appendix:beamlets_selector}). As we show in Appendix~\ref{appendix:theory}
Eq.~\ref{eq10}, the interference of two beamlets that are separated by $\pm x_1$ from the axis results in a pattern that has a particular structure. If two beamlets have no delay between them (no PFT) the pattern is symmetric. However, in the case of a delay of $2\tau_1$ between them (such that the PFT=$\tau_1/x_1$) the internal structure of the fringe is shifted by $\xi_1=\tau_1cf/x_1$ with respect to the main envelope. Therefore to evaluate the PFT it is sufficient to determine the shift of the internal fringe peak $\xi_1$ in the far field. From the shift, we extract the value of the PFT: $\mbox{PFT}=\xi_1/cf$, where $c$ is speed of light in vacuum and $f$ is the focal distance of the focusing optics.
 
\begin{figure}[t]
		\centering
		\includegraphics[width=10cm]{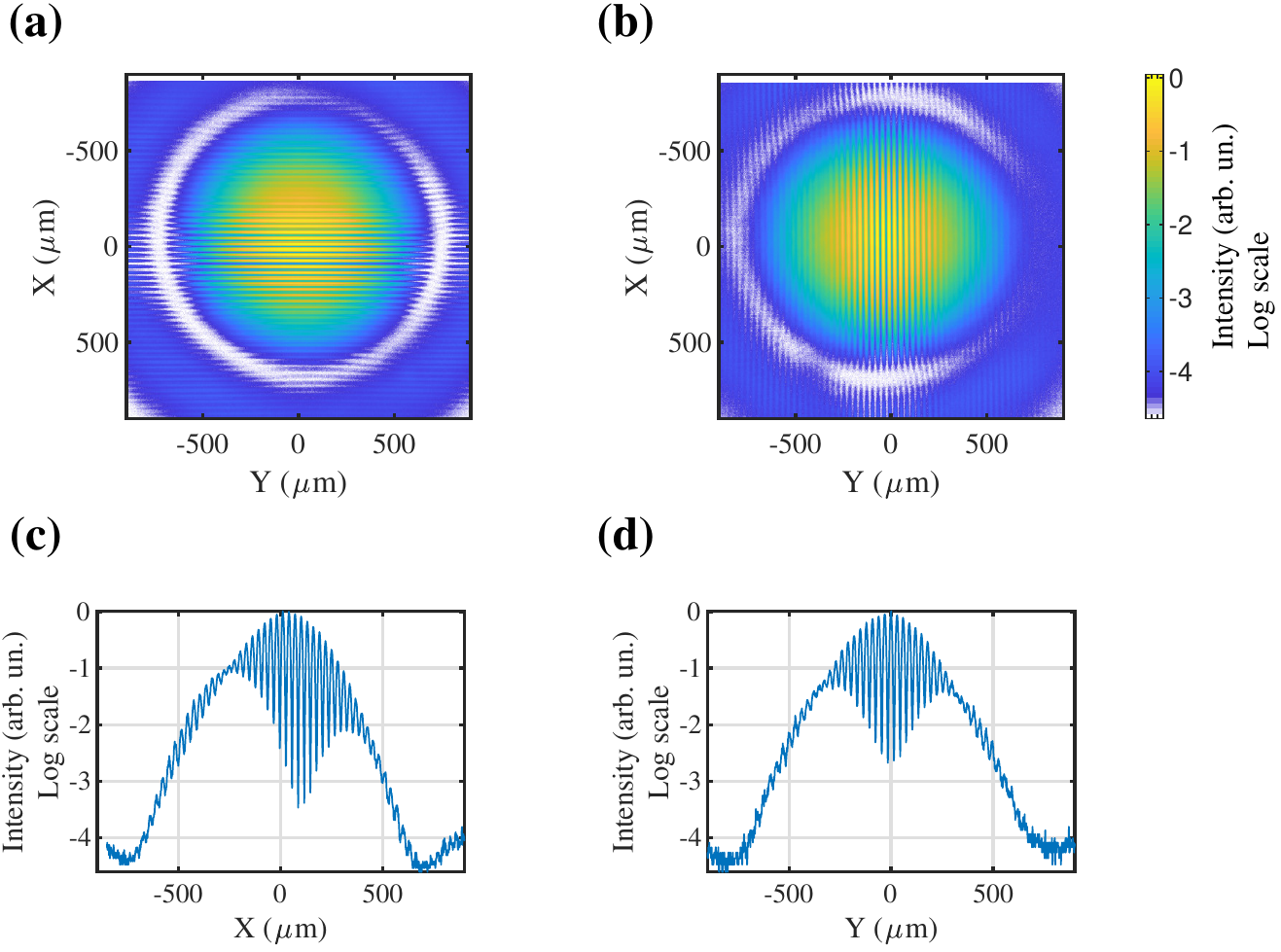}
		\caption{\label{fig:PFT_tool} \small Beamlets interference patterns used to evaluate PFT of the beam along X-axis (a) and Y-axis (b) and their corresponding cross-sections (c) and (d).}
\end{figure}

 We evaluate PFT for the corrected case and choose beamlets which are located close to the beam's edge ($2|x_1| = 54$ mm), symmetrically from the beam center. The interference patterns for the Y and X axes, shown in logarithmic scale, are depicted in Fig.~\ref{fig:PFT_tool} (a-b) and their corresponding cross-sections are in (c-d). It can be seen that the internal fringe envelope in X-axis is shifted by approximately 90 \textmu m, while the fringe pattern along the Y-axis is nearly symmetric. This corresponds to |PFT$_x$|=0.15 fs/mm and |PFT$_y$|=0 which in a good agreement with values obtained in the scan (see the second column of the Table~\ref{table:PFT_PFC}). In addition, the transverse position of the doublet affects the PFT of the beam; we placed it at a transverse position inside the beam expander such that it closely matched the PFT of the uncorrected case. 

In the PFC case, we used the same beamlet selector used in the robust method, which selects one central and one off axis beamlet. Then, by symmetrizing the fringe pattern, similarly to the PFT tool alignment, we minimized PFC by optimizing the longitudinal doublet position inside the beam expander. However, in the case of PFC, this measurement is more sensitive to the misalignment of the mask with respect to the beam, and thus careful alignment is required.

\section{\label{sec:concl}Conclusions and outlook}
In this paper, we presented a novel STC measurement technique based on beamlet cross-correlation that we used to characterize the spatiotemporal pulse front of our 100 TW laser. We found that our laser pulse front is dominated by PFC, which reduces the peak intensity at the focus. A specially designed doublet was shown to almost completely suppress the PFC. In addition, we introduced fast, one-shot pulse-front measurement tools based on the beamlet's interference in the far-field. With the aid of these tools, we aligned the compressor to minimize PFT in x-y and found the optimal position of the corrective doublet for suppressing the PFC.

The advantage of the proposed method is its simplicity: it requires very little specific equipment and is based on a standard "focal spot" diagnostic. As we show with our analytical model, the method has straightforward post-processing, which can be easily implemented from scratch. In addition, the method is not intrusive and is easily scalable for various beam sizes, and can thus be implemented in nearly any optical setup. 

The usage of this method in temporally comparing beamlets along the focal line allows for the measurement of the group velocity of the energy deposition of optics with extended focal depths \cite{Smartsev_19,oubrerie_2021}. 

\section*{\label{sec:ack}Acknowledgments}
The authors would like to thank Guy Han, Gennady Smartsev for mechanical design, Dekel Raanan, Omer Kneller, Mark Baevsky and Louis Daniault for fruitful discussions.

\section*{Funding}
This work was supported by The Schwartz/Reisman Center for Intense Laser Physics, by a research grant from the Benoziyo Endowment Fund for the Advancement of Science, by the Israel Science Foundation, Minerva, Wolfson Foundation, the Schilling Foundation, R. Lapon, Dita and Yehuda Bronicki, and by the Helmholtz association.

\section*{Disclosures}
The authors declare no conflicts of interest.

\section*{Data availability}
Data underlying the results presented in this paper are not publicly available at this time but may be obtained from the authors upon reasonable request.

\section*{}
\authormark{$\dagger$}These authors contributed equally to this work.

\section{Appendix}\label{appendix}
\subsection{Theory} \label{appendix:theory}
The analytical derivation for beamlet cross-correlation is derived in 1D for simplicity. We describe a system in which an ultra-short beam with STC impinges on a perforated mask with two holes, one in the center and another in position $x_1$. For simplicity, the beamlets are assumed to be Gaussian in space, with width $\sigma_0$. The complex spectral amplitude of the two beamlets after the mask is:
\begin{equation} \label{eq1}
A(x,\omega)=A_0 (\omega)\exp \left( -\frac{x^2}{2\sigma_0^2}\right)
+A_1(\omega)\exp\left(-\frac{-(x-x_1 )^2}{2\sigma_0^2}\right)
\end{equation}

We assume that intensity and STC are slowly varying along the beam and, therefore, the spectral and spatial dependencies of each beamlet are separable. The spectral dependence of the beamlets is described by a Gaussian spectrum centered around $\omega_c$ with a bandwidth $c_B$. The ‘global’ spectral phase which is the same for both beamlets is omitted because it will be canceled out in cross-correlation. We consider the relative linear spectral phase, which, in the time domain, is the relative delay between the beamlets and is labeled as $\tau_1$. The spectral dependencies of the beamlets are:

\begin{equation} \label{eq2}
A_0(\omega)=\exp \left( -\frac{(\omega-\omega_c)^2}{2c_B^2}\right)
\end{equation}

\begin{equation} \label{eq3}
A_1(\omega)=RA_0(\omega)\exp \left( i\tau_1(\omega-\omega_c)\right)\exp \left( i\tau_1(\omega_c)\right)
\end{equation}

The $i\tau_1(\omega-\omega_c)$ is the phase which corresponds to a group delay between the beamlets, where $i\tau_1(\omega_c)$ is the constant phase difference between the beamlets. 
$R=|A_1(\omega)|/|A_0(\omega)|$ is the relative amplitude, which accounts for the fact that the beamlets are generated from a beam which has a non-uniform spatial profile. We omit, for simplicity, specifying the complex conjugate terms needed for accurate mathematical description of the pulses in time and frequency.

To calculate the complex spectral field of the beamlet pair on the focal plane we use the Fraunhofer integral \cite{Goodman2005} 

\begin{equation} \label{eq4}
A_f(\xi,\omega)=\int \omega\frac{\exp(ikf)}{i2\pi c f}\exp\left(ik\frac{\xi^2}{2f}\right)A(x,\omega)\exp\left(-ik\frac{x\xi}{f}\right) \,dx
\end{equation}

where $k=\omega/c$ is the k-vector, f is the focal distance of the focusing optics, and $x,\xi$ are the coordinates in the near and far fields respectively. The spatial integral over the $x$ coordinate reduces to a Gaussian (Fourier transform of a Gaussian is a Gaussian). The $x_1$ shift of the off-centered beamlet corresponds to a linear phase in the far field. For simplicity, we assume a narrow spectral bandwidth $c_B/\omega_c<<1$ so that the spatial dependence is dictated by a central wavelength $k\rightarrow k_c$:

\begin{equation} \label{eq5}
A_0:\int \exp\left(\frac{x^2}{2\sigma_0^2}\right)\exp\left(-ik\frac{x\xi}{f}\right) \,dx \propto \exp\left(-\frac{k_c^2\sigma_0^2}{2f^2}\xi^2\right)
\end{equation}

\begin{equation} \label{eq6}
A_1:\int \exp\left(\frac{(x-x_1)^2}{2\sigma_0^2}\right)\exp\left(-ik\frac{x\xi}{f}\right) \,dx \propto \exp\left(-\frac{k_c^2\sigma_0^2}{2f^2}\xi^2\right)\exp\left(-\frac{ix_1k_c}{f}\xi\right)
\end{equation}

The first terms in the Fraunhofer integral can be omitted since they don't affect the final result: $\omega / 2\pi c f$ is constant under the narrow bandwidth approximation, and the phase terms $\exp(ikf)/i$ and $\exp(ik\xi^2/2f)$ disappear when moving from field to intensity.
The final expression is

\begin{equation} \label{eq7}
A_f(\xi,\omega)\propto \exp \left( -\frac{(\omega-\omega_c)^2}{2c_B^2}\right)
\left[1+R \exp\left(i\left[(\omega-\omega_c)\tau_1+\omega_c\tau_1-\frac{x_1k}{f}\xi\right]\right)\right]\exp\left(-\frac{k_c^2\sigma_0^2}{2f^2}\xi^2\right)
\end{equation}

Therefore the complex field in space-time in the far field is
\begin{equation} \label{eq8}
A_f(\xi,t)=\int \exp(i\omega t)A_f(\xi,\omega) \,d\omega
\end{equation}

It can be proved that, up to constants, the far field intensity integrated over time is

\begin{align}\label{eq9}
I_f(\xi)&=\int |A_f(\xi,t)|^2 \,dt =\nonumber\\
&\exp\left(-\frac{\sigma_0^2 \omega_c^2}{c^2 f^2}\xi^2\right)
\left[\frac{1+R^2}{2}+R \exp\left(-\frac{c_B^2}{4}\left(\tau_1+\xi\frac{x_1}{c f}\right)^2\right)\cos\left(\omega_c\left[\tau_1+\xi \frac{x_1}{c f}\right] \right)\right]
\end{align}

A closer look at the final expression shows that it corresponds to a localized fringe pattern with an internal structure. The first exponent determines the overall envelope, which is Gaussian. The internal structure consists of fringes with spatially varying contrast. The fringe spacing is determined by the argument inside of the cosine: the k-vector difference between the test and reference beamlets. The fringe contrast shape is determined by the second Gaussian exponent which has the delay $\tau_1$ in the argument.

It is easy to extend this result to obtain the equation of far field intensity (Eq.~\ref{eq9}) for two beamlets which are separated symmetrically from the optical axis at $-x_1$ and $x_1$, having delay $-\tau_1$ and $\tau_1$ (as is the case with the "PFT tool"):

\begin{align}\label{eq10}
I_{fsym}(\xi)&=\int |A_{fsym}(\xi,t)|^2 \,dt =\nonumber\\
&\exp\left(-\frac{\sigma_0^2 \omega_c^2}{c^2 f^2}\xi^2\right)
\left[\frac{1+R^2}{2}+R \exp\left(-c_B^2\left(\tau_1+\xi\frac{x_1}{c f}\right)^2\right)\cos\left(2 \omega_c\left[\tau_1+\xi \frac{x_1}{c f}\right] \right)\right]
\end{align}

It is convenient to perform a spatial Fourier transform of the fringe pattern intensity $I_f(\xi)$ over the spatial coordinate $\xi$. The reciprocal coordinate (or spatial frequency) is labeled $k_\xi$

\begin{equation} \label{eq11}
\widetilde{I_f}(k_\xi)=\mathfrak{F}(I_f(\xi))
\end{equation}

As expected the Fourier transform of the intensity fringe pattern (which has no phase by definition) consists of a DC peak and two sidebands. It could be shown that the normalized spatial Fourier transform of the fringe pattern is:

\begin{equation} \label{eq12}
\widetilde{I_f}(k_\xi)=\exp(a_{DC})+\Gamma_{SB}[\exp(a_{SB1})+\exp(a_{SB2})]
\end{equation}

where
\begin{equation} \label{eq13}
a_{DC}=-\frac{c^2f^2k_\xi ^2}{4\sigma_0^2\omega_c^2}
\end{equation}

\begin{equation} \label{eq14}
a_{SB1,2}=-\frac{[c_B\tau_1 \sigma_0 \omega_c]^2+(cfk_{\xi}\pm \omega_c x_1)^2+i\tau_1[x_1c_B^2 c f k_{\xi}\mp 4\sigma_0^2 \omega_c^3]}{4\sigma_0^2\omega_c^2+c_B^2 x_1^2}
\end{equation}

As we can see, the amount of group delay $\tau_1$ between the beamlets is present in the argument of the exponent of the sidebands, affecting the amplitude and the phase.

The relative amplitude of the side-bands is

\begin{equation} \label{eq15}
\Gamma_{SB}=\frac{R}{(1+R^2)\sqrt{1+\frac{c_B^2 x_1^2}{4\sigma_0^2 \omega_c^2}}}
\end{equation}

To invert the relation, we solve the quadratic equation and get

\begin{equation} \label{eq16}
R=\frac{1-\sqrt{1-\Gamma_{SB}^2\left(4+\frac{c_B^2x_1^2}{\sigma_0^2\omega_c^2}\right)}}{\Gamma_{SB}\sqrt{4+\frac{c_B^2x_1^2}{\sigma_0^2\omega_c^2}}}
\end{equation}

If we look at the maxima of the sidebands, located at $|cfk_\xi\pm \omega_c x_1|=0$, they have a Gaussian dependence on the delay. The delay $\tau_1$ is composed of the delay from the STC and from the delay imposed by the piezo actuator. In other words, when $\tau_1$ is zero, the two beamlets are synchronized. Therefore, scanning the delay with a piezo and finding the maximum reveals the STC delay. The Gaussian dependence is shown below:

\begin{equation} \label{eq17}
\exp\left(-\frac{1}{2\left(\frac{2}{c_B^2}+\frac{x_1^2}{2\sigma_0^2 \omega_c^2}\right)}\tau_1^2\right)=\exp\left(-\frac{1}{2\sigma_{fc}^2}\tau_1^2\right)
\end{equation}
Under this Guassian approximation, when the temporal width of the two beamlets is the same, their cross-correlation at $x_1=0$ - which is equivalent to the autocorrelation - is twice as large as the intensity of the transform limited pulses $\sigma_{TL}^2=1/2c_B^2$, which agrees with the field autocorrelation theorem \cite{UltrashortBook}. Per the same theorem, our measurement is insensitive to a common phase between the two beamlets. Thus, this method can't measure global spectral phase such as global GDD.

However, relative spectral phase (or relative spectral narrowing) which broadens the pulse can be estimated. Assume relative GDD $\beta_r(x_1)$, which broadens only one of the beamlets. It can be shown that the beamlet cross-correlation takes the form

\begin{equation} \label{eq18}
\exp\left(-\frac{1}{2\sigma_{bc}^2}\tau_1^2\right)=\exp\left(-\frac{1}{2\left[\frac{2}{c_B^2}+\frac{x_1^2}{2\sigma_0^2\omega_c^2}+\frac{c_B^2\beta_r(x_1)^2}{2}\right]}\tau_1^2\right)
\end{equation}
In order to get the pulse intensity in time we raise the cross-correlation to the fourth power so that we get the same width as for a transform limited intensity pulse:
\begin{equation} \label{eq19}
\exp\left(-\frac{1}{2\sigma_{bc}^2}\tau_1^2\right)^4=\exp\left(-\frac{1}{2\sigma_{bc4}^2}\tau_1^2\right)
\end{equation}

\begin{equation} \label{app:eq20}
\sigma_{bc4}^2=\frac{1}{2c_B^2}+\frac{x_1^2}{8\sigma_0^2\omega_c^2}+\frac{c_B^2\beta_r(x_1)^2}{8}
\end{equation}

The width squared of the beamlet cross-correlation raised to the fourth power, $\sigma_{bc4}^2$, consists of three terms: the transform limited $\sigma^2 = 1/2c^2_B$, a geometric broadening factor and a relative GDD broadening factor.

\subsection{Beamlets cross-correlation spectrum and deconvolution factor} \label{appendix:BCC_spectrum}

\begin{figure}[ht]
		\centering
		\includegraphics[width=5cm]{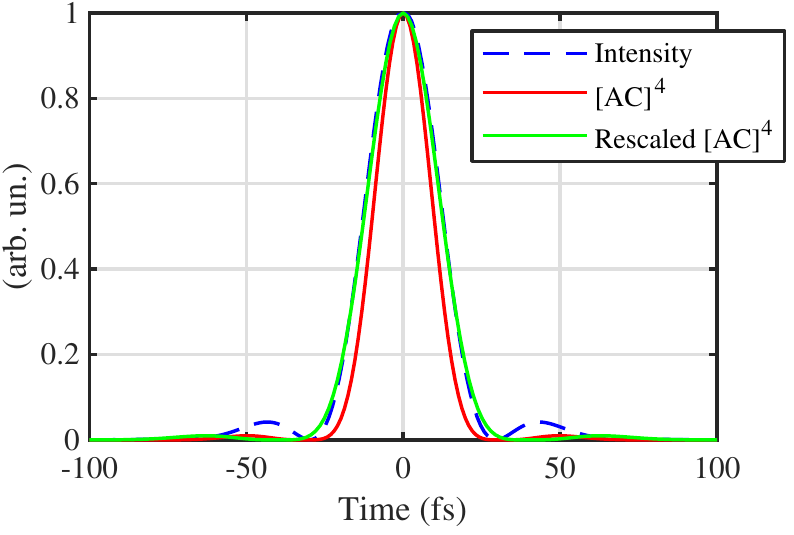}
		\caption{\label{fig:deconv_fact} \small {Field auto-correlation raised to the fourth power compared to intensity: (dashed blue) Wizzler measured intensity, (red) auto-correlation raised to the power of four of the field derived from Wizzler measured intensity, (green solid) re-scaled autocorrelation raised to the power of four which matches the measured intensity.}}
\end{figure}

\begin{figure}[ht]
		\centering
		\includegraphics[width=10cm]{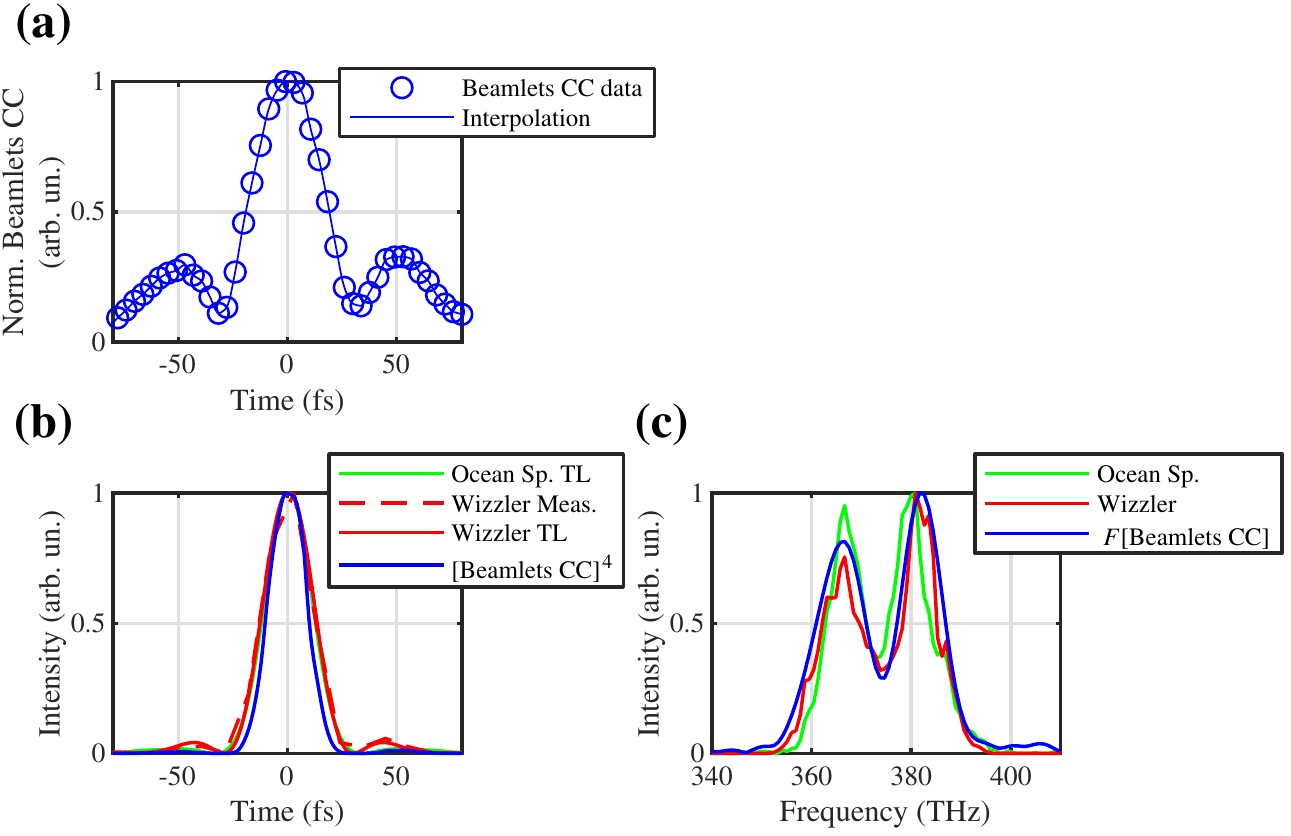}
		\caption{\label{fig:spectra} \small {Temporal and spectral intensities. \textbf{(a)} Normalized beamlet cross-correlation envelope data (circles) and interpolating curve (solid). \textbf{(b)} Beamlet cross-correlation envelope raised to the fourth power (solid blue) plotted with Wizzler measured (dashed red) and transform-limited (TL) (solid red) temporal intensities, and fiber-coupled spectrometer (Ocean Optics) TL pulse (solid green). \textbf{(c)} Spectral intensity from Fourier transform of beamlet cross-correlation (solid blue), with Wizzler (solid red) and fiber-coupled spectrometer (solid green) spectra.}}
\end{figure}

In our experiment, the spectrum is a double peak rather than Gaussian (see Fig. \ref{fig:spectra} (c)). Therefore, the beamlet cross-correlation width $\sigma_{bc4}$ is not exactly the same as the intensity temporal width. We calculate the deconvolution factor which properly scales the beamlet cross-correlation width $\sigma_{bc4}$ with the pulse temporal width. For this we cross-correlate the Wizzler measured field with itself (autocorrelation), raise the result to the fourth power and compare it with the field's intensity width (see Fig.~\ref{fig:deconv_fact}). We get that for our spectrum, $\sigma_{bc4}$ is smaller by a factor of 1.27 compared to the intensity width $\sigma$.

\subsection{Delay mirror} \label{appendix:delay_mirror}
\begin{figure}[t]
		\centering
		\includegraphics[width=13cm]{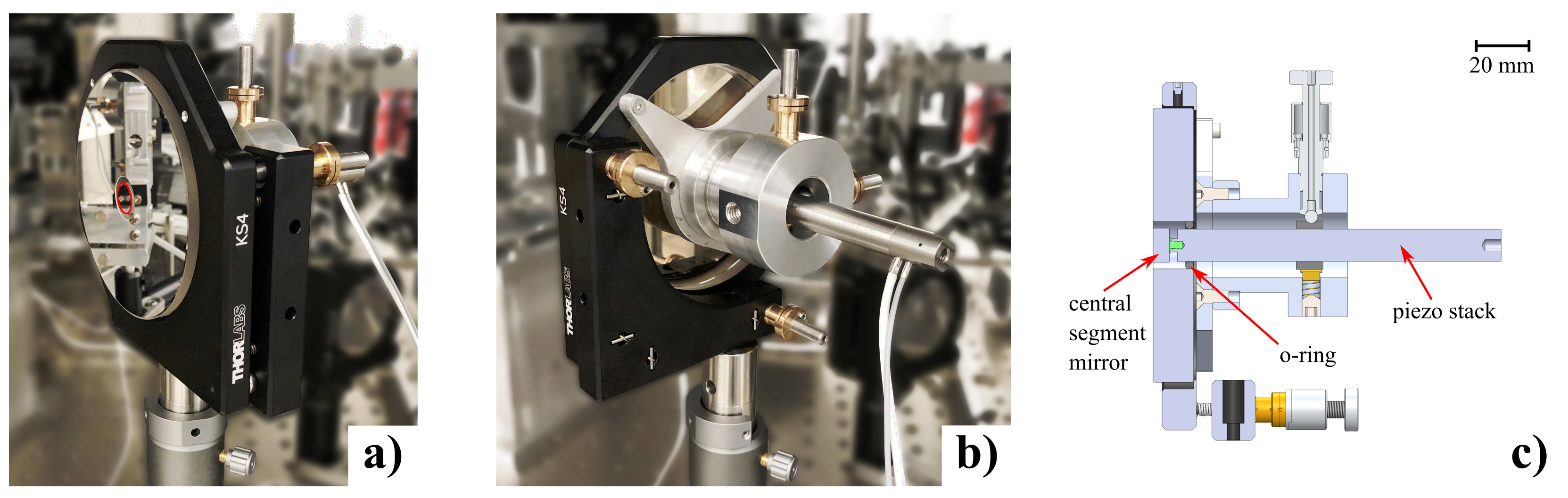}
		\caption{\label{fig:SegMirr} \small {Segmented delay mirror assembly. \textbf{(a)} Front view: the central segment (marked in red) position is controlled with the piezo stack. \textbf{(b)} Rear view: The mirror has independent angle control for the central segment and both mirrors together.}}
\end{figure}

The segmented delay mirror assembly is depicted in Fig.~\ref{fig:SegMirr}. The assembly is based on a common 4" mount (Thorlabs KS4) which holds the main 4" diameter 15 mm thick plane mirror. The mirror has a central hole with a diameter of 18 mm. The hole, with the help of an o-ring, serves as an attachment for the piezo stack. This assures that the axis of tilt is as close as possible to the surface of the large mirror Fig.~\ref{fig:SegMirr} (c). The homemade holder for the piezo actuator allowed for independent control of the central segment mirror with respect to the large mirror. For the central mirror, we used a 1/2" diameter mirror (Thorlabs PF05-03-P01). We used a piezo stack (PI-P-840) which has nanometric precision over a 90 $\mu m$ range. The piezo stack has a position sensor and is operated in a closed-loop. The compact size of the assembly allows for easy and non-intrusive installation in nearly any experimental setup.

\begin{figure}[t]
		\centering
		\includegraphics[width=2cm]{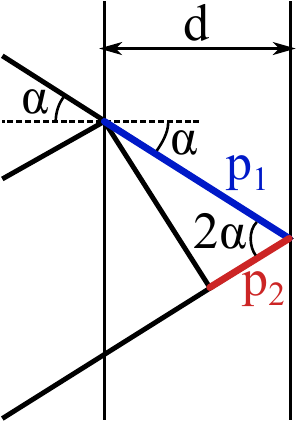}
		\caption{\label{fig:delay_angle} \small {Delay imposed on optical path for the non-zero reflected ray.}}
\end{figure}

For the case of non-zero reflection from the central mirror, the path difference depends on the angle. Consider a ray which is impinging on the mirror at an angle $\alpha$ and the mirror is shifted by a distance of $d$. The total delay difference $d_{tot}$ as can be seen in Fig.~\ref{fig:delay_angle} is composed of $p_1+p_2$
\begin{equation} \label{eq:app:delay}
d_{tot}=\frac{d}{\cos (\alpha)}(1+\sin (90-2\alpha))
\end{equation}

\subsection{Beamlet selector} \label{appendix:beamlets_selector}
\begin{figure}[t]
		\centering
		\includegraphics[width=10cm]{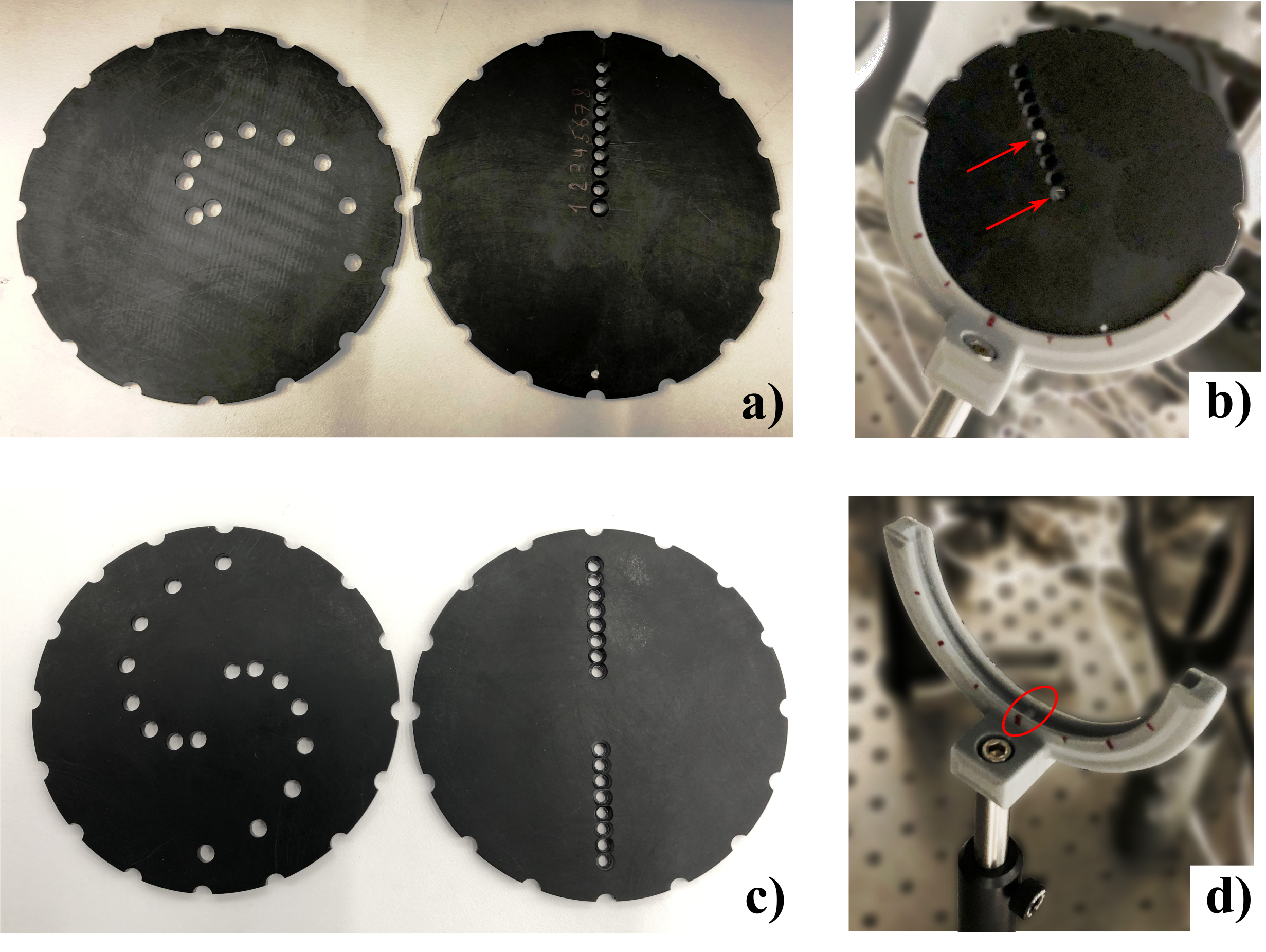}
		\caption{\label{fig:BeamSel} \small {Beamlets selector and "PFT tool" masks. \textbf{a)} Beamlet selector mask selects on-axis and off-axis beamlets. \textbf{(b)} A pair of beamlets selected for a specific angle and radius. \textbf{(c)} "PFT tool" mask which selects two off-axis beamlets. \textbf{(d)} 3D printed plastic mount holds the pair of disks. The special bump (marked in red) snaps the disks to chosen position. }}
\end{figure}

The beamlet selector is presented in Fig.~\ref{fig:BeamSel} and consists of two main parts: aperture disks and a holder. The disks have a 100 mm diameter and are laser cut from 3 mm Delrin plates (Fig.~\ref{fig:BeamSel}). Both disks were sanded to roughen the surface to achieve diffusive reflection of the blocked laser beam. One disk has radially distributed holes of 2.65 mm diameter (right hand side on Fig.~\ref{fig:BeamSel}(a)); the holes were beveled to increase angular acceptance. The second disk consists of spirally oriented holes with a diameter of 5 mm (left hand side on Fig.~\ref{fig:BeamSel}(a)). When the disks are superimposed, the central hole stays open (reference beamlet), and the relative orientation of the disks allows radial selection of the second opened hole (test beamlet). The orientation of both disks together sets the angle of the test beamlet. The disk holder is 3D printed using Polylactic Acid (PLA) and holds the two disks securely (Fig.~\ref{fig:BeamSel}(b)). The holder has a special bump (marked in red) that snaps to the notches made at the edges of the disks (Fig.~\ref{fig:BeamSel}(d)). This allows for precise selection of test beamlet for a discrete set of radii and angles, as shown in Fig.~\ref{fig:BeamSel}(b).

\begin{figure}[t]
		\centering
		\includegraphics[width=13cm]{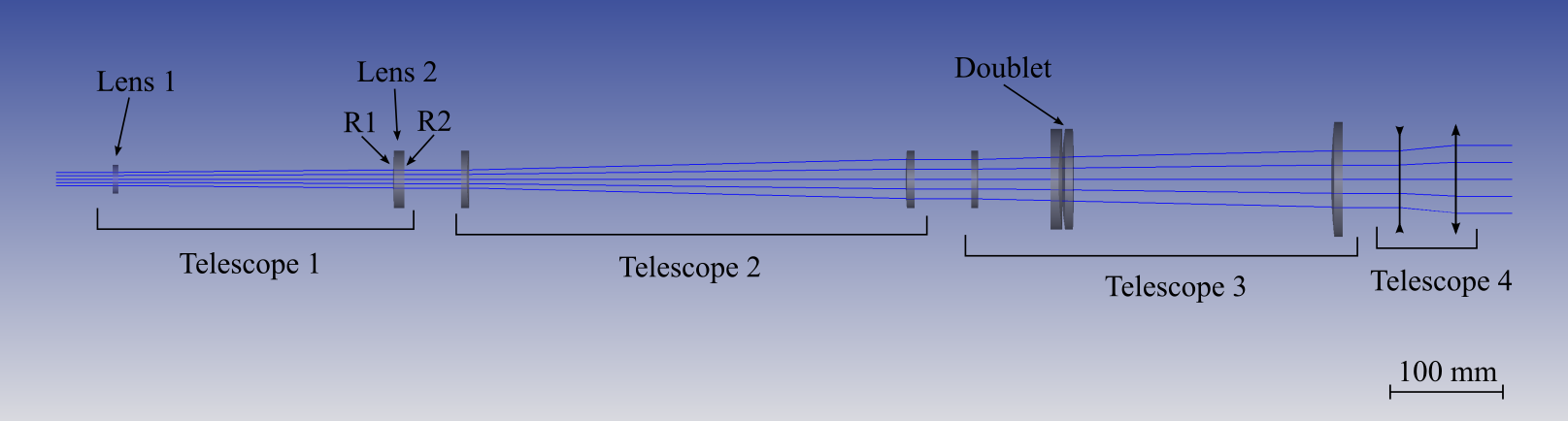}
		\caption{\label{fig:Zemax_laser_chain} \small {Optical system used for the laser chain STC simulation. The system consists of 4 telescopes (beam expanders) and a corrective doublet which is located inside of telescope 3. The numbering of the telescopes, lenses and their radii are from left to right. Telescopes 1-3 are refractive and telescope 4 is an ideal telescope which has no dispersion.}}
\end{figure}

\begin{table}[ht]

\renewcommand{\arraystretch}{2}
\footnotesize
\centering
\begin{tabular}{ |P{1.4cm}||P{1.9cm}|P{0.9cm}|P{0.9cm}|P{0.9cm}| }

 \hline
  &(Distance units are mm) &Telescope 1&Telescope 2&Telescope 3\\
 \hline\hline
 &Magnification&	1.37&	2.20&	1.43\\
 \hline
 Lens 1 (F. silica)&R1&	0&	0&	5\\
 &R2&	305&	150&	330\\
& Thickness&	5&	5&	5\\
 \hline
&Lens 1-2 separation&	250	&395&	320\\
 \hline
Lens 2 (F. silica) &R1&	420.2	&325.5&	472.4\\
&R2&	0&	0&	0\\
&Thickness&	10&	8	&10\\
 \hline
\end{tabular}
\caption{\label{table:telescopes} \small {Technical details of the refractive telescopes}}
\end{table}

\subsection{Laser chain STC simulation} \label{appendix:laser_chain}
We simulated the STCs of the laser chain at WIS and the correction of PFT by a special doublet using the Zemax OpticStudio ray-tracing software. We considered only three refractive telescopes, even though the WIS laser consists of more beam expansion stages. This choice was motivated by the fact that the PFC induced by the beam expander (assuming the same material for both lenses) depends on the beam size and focusing properties of the lenses \cite{Bor_telescope}. Since all our telescopes have similar focusing properties, the PFC is proportional to the beam size. Therefore, the earlier stages, in which the beam radius is small, play a negligible effect in the STC and are thus ignored. In the last beam expander, the beam size is largest and, therefore, most of the STC is accumulated there. The system is shown in Fig.~\ref{fig:Zemax_laser_chain}. The simulation begins with a broad spectrum beam with perfect spatial phase and no STC which is gradually expanded by refractive Galilean telescopes – labeled telescope 1-3 – and accumulates STC with minor spatial aberrations. STC correction is applied via a specially designed doublet, inserted inside telescope 3. Precise element-wise details about the telescopes are depicted in Table~\ref{table:telescopes}, and the corrective doublet is described in Table~\ref{table:doublet}. In the real system, the beam propagates around 21 meters before reaching the focusing parabola. Since the beam diverges slightly, this results in an increase of its size by a factor of 1.2. We have verified by ray-tracing that the flat glass window and compressor (assumed aligned) introduce negligible STC to the beam; therefore, we did not consider them here. Also, we did not consider amplification crystals between the telescopes. We simulated the divergence in free space as "Telescope 4" - a perfect dispersionless beam expander. The induced PFC will be overestimated compared to the measured one if the dispersionless beam size growth is not included in the simulation.

\begin{table}[ht]
\renewcommand{\arraystretch}{2}
\footnotesize
\centering
\begin{tabular}{ |P{2cm}||P{1cm}|P{1cm}| }

 \hline
( Distance units are mm)&	Lens 1	&Lens 2\\
 \hline\hline
 Glass (CDGM)&	H-ZF1&	H-K9L\\
\hline
 R1&	0&	306\\
R2&	306&	-1331\\
 Thickness&	7&	11\\
\hline
 Lens 1-2 separation&\multicolumn{2}{c|}{ 2.55} \\
 \hline
\end{tabular}
\caption{\label{table:doublet} \small {Technical details of the correction doublet.}}
\end{table}

\begin{figure}[t]
		\centering
		\includegraphics[width=13cm]{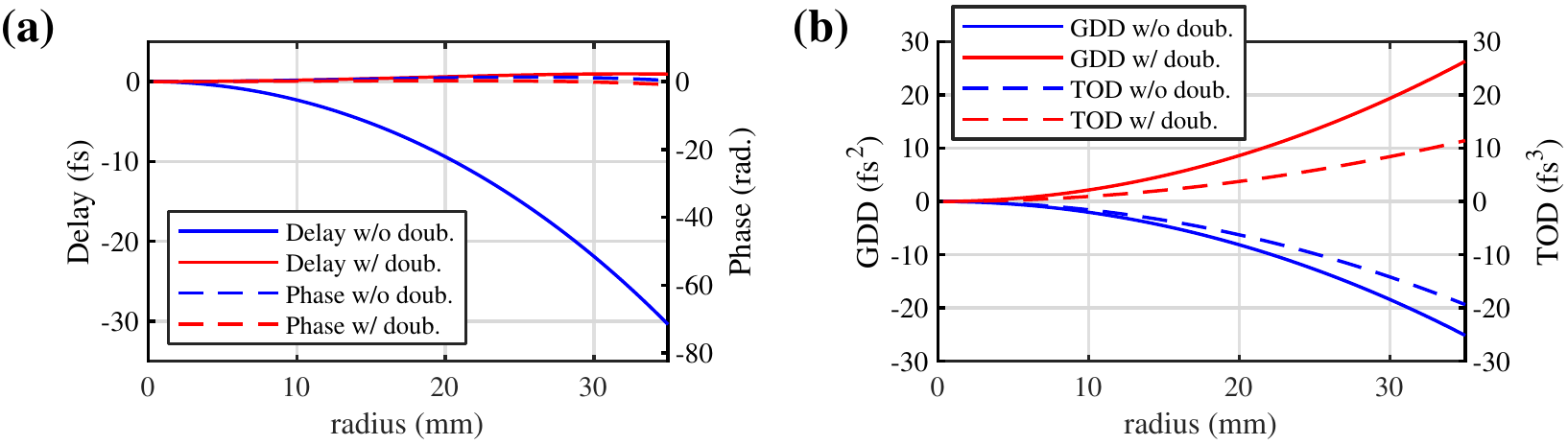}
		\caption{\label{fig:zemax_stc} \small {Simulated relative values of dispersion along the beam. (a) Group delay (solid) and phase (dashed) for 800 nm with (blue) and without (red) corrective doublet. (b) GDD (solid) and TOD (dashed) with (blue) and without (red) corrective doublet.}}
\end{figure}

We modified the macro\cite{OShea_Appl_Opt_06} for Zemax OpticStudio to calculate pulse group delay, group velocity dispersion, and third-order spectral phase across the beam. For each position in the beam, the macro traces rays in the vicinity of the central wavelength (800 nm in our case) and calculates the optical path. Then, using derivatives, it approximates relevant dispersion parameters. We consider only a radially symmetric case since we assume perfect alignment and optics. The results of the simulation are depicted in Fig.~\ref{fig:zemax_stc}. The group delay along the beam is shown in Fig.~\ref{fig:zemax_stc}(a). As can be seen, without the corrective doublet, the simulated beam has pulse front curvature (the central part of the beam is delayed more). When the corrective doublet is present, the curvature is suppressed. By fitting to a parabola we get the value of the PFC: PFC=-0.0243 (fs/mm$^2$) and PFC=0.0011 (fs/mm$^2$) for the uncorrected and the corrected cases respectively. The doublet position inside of telescope 3 mimicked the position it had in the experiment described in this paper. The doublet introduces a small focusing term which is corrected by adjusting the position of the focal plane of the final focusing parabola (several mm with 2 m focal length parabola). The doublet introduces negligible aberrations at all positions inside the telescope; thus, the beam remains almost diffraction-limited.

In the same figure, we show the accumulated phase difference, as a function of radius, of a 800 nm monochromatic wave propagating through this optical setup. We can see that the phase difference is close to zero. Assuming an 800 nm wave propagating in vacuum, the value of the phase and the delay differ only by a numerical factor and thus, the figure shows both the phase and the delay as a function of radius. The quadratic focusing term from the prior optics in the simulation is eliminated by adjusting the last lens of telescope 4. In the real system, the curved spatial front of the diverging beam is compensated by focal distance adjustment of the focusing mirror. In Fig.~\ref{fig:zemax_stc} (b) higher orders of dispersion as a function of radius are presented. As seen in both the uncorrected and corrected cases, the simulated beam gets some relative group delay dispersion (GDD) and third order dispersion (TOD) which broadens the pulse but is negligible in our case. This stems from the fact that the beam is gradually expanded through the telescopes, which have the same dispersion (same glass) but radially differing thickness of glass and angle of incidence. This causes the central part of the beam to propagate through more glass and accumulate more group delay and more relative GDD and TOD than the outer parts of the beam. In the corrected case, special glasses and curvatures are used in the doublet to cancel the PFC; however, the relative GDD and TOD are slightly over-corrected. The relative dispersion values, related to r=0 mm, are presented in Table~\ref{table:global_gdd} along with the global dispersion values at r=0 mm.

\begin{table}[ht]
\renewcommand{\arraystretch}{2}
\footnotesize
\centering
\begin{tabular}{ |P{1.4cm}||P{1.4cm}|P{1cm}|P{1cm}| }

 \hline
 &Group Delay(ps)& GDD(fs$^2$) &TOD(fs$^3$)\\
 \hline\hline
No doublet &4357& 1545& 1156\\
 \hline
With doublet&4392& 2923& 2069\\
 \hline
\end{tabular}
\caption{\label{table:global_gdd} \small {Global dispersion values at r=0 mm.}}
\end{table}

\bibliography{STC_BIB}

\end{document}